\documentclass[journal]{IEEEtran}
\IEEEoverridecommandlockouts
\usepackage{cite}
\usepackage{amsmath,amssymb,amsfonts,mathtools,bm}
\usepackage{amsthm}
\usepackage{algorithm}
\usepackage{algpseudocode}
\usepackage{graphicx}
\usepackage[table]{xcolor}
\usepackage{booktabs}
\usepackage{array}
\usepackage{multirow}
\usepackage{microtype}
\usepackage{url}
\usepackage{hyperref}
\usepackage{stfloats}
\usepackage{tikz}


\newtheorem{problem}{Problem}
\newcommand{\ind}{\mathbb{1}}

\definecolor{lime}{HTML}{A6CE39}
\DeclareRobustCommand{\orcidicon}{%
  \begin{tikzpicture}[baseline=-0.55ex]
  \draw[lime,fill=lime] (0,0) circle [radius=0.16]
  node[white] {{\fontfamily{qag}\selectfont\tiny ID}};
  \end{tikzpicture}\hspace{-1.2mm}}
\newcommand{\orcidauthorA}{0000-0003-1439-4875}
\newcommand{\orcidauthorC}{0000-0001-7907-2071}
\newcommand{\orcidauthorH}{0000-0002-3136-788X}
\newcommand{\orcidauthorF}{0000-0002-5727-2432}
\newcommand{\orcidauthorE}{0000-0002-4403-9893}
\newcommand{\orcidA}{\href{https://orcid.org/\orcidauthorA}{\orcidicon}}
\newcommand{\orcidC}{\href{https://orcid.org/\orcidauthorC}{\orcidicon}}
\newcommand{\orcidH}{\href{https://orcid.org/\orcidauthorH}{\orcidicon}}
\newcommand{\orcidF}{\href{https://orcid.org/\orcidauthorF}{\orcidicon}}
\newcommand{\orcidE}{\href{https://orcid.org/\orcidauthorE}{\orcidicon}}
\allowdisplaybreaks
\renewcommand{\baselinestretch}{1}

\begin{document}

\title{{\color{black}SateBeam: }Beam-Aware Radio Map Estimation With Physics-Consistent Parametric Modeling for Unknown Multiple Satellites}
\author{
Xiucheng Wang,\orcidA{}~\IEEEmembership{Student Member,~IEEE,}
Nan Cheng,\orcidC{}~\IEEEmembership{Senior Member,~IEEE,}
Zhisheng Yin,\orcidH{}~\IEEEmembership{Member,~IEEE,}
Conghao Zhou,\orcidF{}~\IEEEmembership{Member,~IEEE,}
and Ruijin Sun,\orcidE{}~\IEEEmembership{Member,~IEEE}
\thanks{This work was supported by the National Key Research and Development Program of China (2024YFB907500).}
\thanks{Xiucheng Wang, Nan Cheng, Zhisheng Yin, Conghao Zhou, and Ruijin Sun are with the State Key Laboratory of ISN and School of Telecommunications Engineering, Xidian University, Xi'an 710071, China (e-mail: xcwang\_1@stu.xidian.edu.cn; dr.nan.cheng@ieee.org; zsyin@xidian.edu.cn; conghao.zhou@ieee.org; sunruijin@xidian.edu.cn). Nan Cheng is the corresponding author.}}
\maketitle

\begin{abstract}
Dense low-Earth-orbit (LEO) networks increasingly reuse spectrum across moving satellites and overlapping beams, making a spatial radio map essential for interference diagnosis and coordination. Constructing such a map from sparse ground power measurements is difficult because the active satellite subset and its cardinality are unknown, candidate footprints can be highly coherent, and beam pointing, width, effective power, Doppler, and background must be inferred jointly. Known ephemerides do more than enumerate visible candidates. Propagated positions determine candidate-specific ranges, viewing geometry, and spatial footprints, while radial velocities induce distinct time-frequency signatures. {\color{black}Their joint spatial--Doppler structure rejects candidates whose geometry or time--frequency signature conflicts with the observations before continuous beam inference. SateBeam therefore screens all candidates through Doppler-resolved nonnegative fitting. GPU-accelerated grid correlations initialize continuous off-grid refinement; a Gamma-likelihood Bayesian information criterion (BIC) selects model cardinality, and backward deletion and exchange repair greedy support errors.} {\color{black}Across paired real-ephemeris scenarios, SateBeam recovers the exact support in every case and reduces independent-query RMSE from $0.655$ to $0.050$ dB relative to orthogonal matching pursuit, with consistent advantages under model mismatch and larger candidate sets.} {\color{black}The code is available at \url{https://github.com/UNIC-Lab/SateBeam}.}
\end{abstract}


\section{Introduction}
Dense LEO constellations create rapidly changing co-channel interference through aggressive spectrum reuse, overlapping spot beams, and multi-constellation operation \cite{giordani2020non,del2019technical,heydarishahreza2024spectrum}. A ground-side radio map (RM) that predicts received power across a region can support interference localization, beam/frequency coordination, spectrum sharing, and operational verification. Unlike a terrestrial coverage map whose transmitters are often fixed and known, a satellite RM must attribute a moving aggregate field to an unknown subset of visible satellites.

This task is challenging for three coupled reasons. First, ephemeris visibility identifies possible satellites, whereas payload illumination and in-band activity must be inferred. Second, beam centers, widths, effective in-band powers, and background are unknown continuous quantities, and different satellites can produce nearly proportional footprints over a finite region. Third, LEO motion changes geometry and Doppler during measurement, and the available observation is averaged power. The resulting inverse problem has an unknown discrete support, nonconvex continuous beam parameters, coherent candidate signatures, and heteroscedastic energy measurements. Reliable satellite-attributed mapping therefore requires identity recovery and spatial prediction within one inference model.

Known ephemerides provide two complementary constraints. Their propagated positions determine range, elevation, off-nadir angle, atmospheric loss, and the spatial radiation footprint. Their radial velocities create candidate-specific Doppler trajectories across time and frequency. {\color{black}Geometry and Doppler jointly narrow the candidate support before continuous spatial fitting, enabling a compact, interpretable beam model over the retained candidates.}

{\color{black}This coupled structure motivates SateBeam for joint active-satellite identification and satellite RM reconstruction. SateBeam first performs all-candidate Doppler nonnegative least-squares (NNLS) screening from subband power spectra. It then uses graphics processing unit (GPU) grid correlations to propose beam initializations, continuously refines the complete trial model before applying a Gamma-likelihood Bayesian information criterion (BIC) rule, and corrects greedy path dependence through backward deletion and one-for-one support exchange.} The forward model explicitly includes effective isotropic radiated power (EIRP), receiver gain, frequency, slant range, atmospheric and polarization loss, Doppler capture, thermal noise, and an estimated background. The nominal beam follows a circular-aperture pattern tied directly to half-power beamwidth (HPBW). The main contributions of this paper are summarized as follows.
\begin{enumerate}
\item {\color{black}We formulate satellite RM construction as a Doppler-resolved averaged-power inverse problem with known ephemerides, an unknown active subset and cardinality, unknown continuous beam parameters, and an explicitly estimated background. The model states which link-budget quantities are known, fitted, or treated as mismatch.}
\item {\color{black}We develop SateBeam, a support-correcting solver that couples Doppler NNLS screening, bounded GPU beam proposals, post-refinement Gamma-BIC selection, backward deletion, and support exchange. The method infers model order directly from averaged-power observations.}
\item {\color{black}We provide complexity and identifiability diagnostics together with reproducible real-ephemeris experiments. SateBeam reaches F1 $1.000$, 100\% exact support, and 0.050 dB query RMSE in the main paired study; it improves over OMP under sidelobe, elliptical-beam, shadowing, and blockage mismatch and scales to $N=64$ candidates and $K=8$ active sources.}
\end{enumerate}

The remainder follows the inference logic. Section~\ref{sec:related} positions the problem within satellite beam management and RM reconstruction. Section~\ref{sec:system} derives the geometry, beam/link-budget, Doppler, and power-observation model. Section~\ref{sec:method} shows how each part of SateBeam follows from that model. Section~\ref{sec:experiments} evaluates the physical assumptions, support recovery, map accuracy, mismatch, and complexity.

\section{Related Work}
\label{sec:related}
\subsection{Satellite Beam and Interference Management}
LEO communication studies analyze constellation geometry, beam placement, power allocation, beam hopping, and interference mitigation under known network control variables \cite{choi2023joint,lei2020beam,hu2020dynamic,lin2022multi,cottatellucci2006interference}. Satellite cognitive-radio and satellite-over-satellite architectures further show that interference observability depends on the radiation interface and network information boundary \cite{sharma2012satellite,lee2000satellite}; standardized NGSO radiation-pattern envelopes provide a complementary reference for antenna-model assumptions \cite{itu1528}. {\color{black}These works optimize known network control variables; SateBeam addresses the complementary inverse task of inferring active satellites and beams from ground power.} We therefore use ephemerides as known sensing structure while leaving payload activity, beam commands, and effective EIRP unknown. The downlink radiation-interface formulation covers both transparent and regenerative payload architectures through their effective in-band patterns.

\subsection{Radio Maps and Satellite Reconstruction}
Terrestrial RM methods include spatial interpolation, Gaussian processes, low-rank completion, physics-aware learning, and generative reconstruction \cite{zeng2021toward,romero2024theoretical,levie2021radiounet,wang2025radiodiff}. Earlier radio-environment-map work established cognitive-network and interference-control uses \cite{YilmazBirkan2013,sato2017kriging}, while Gaussian-process and nearest-neighbor variants address sparse spatial sampling \cite{wang2020indoor,zhang2022k}. Public prediction challenges and learning-based estimators now cover convolutional reconstruction, active sampling, geometry awareness, graph models, and physics-informed representations \cite{yapar2023first,yapar2025sampling,chaves2023deeprem,10735108,10745208,li2024radiogat,jia2025physics}. Recent high-capacity variants explore diffusion and flow matching, three-dimensional and multipath-aware benchmarks, Transformers, super-resolution, and task-driven generation \cite{wang2024radiodiff,11282987,11083758,11278649,jia2025rmdm,jia2025radioflow,wang2025iradiodiff,li2025rmtransformer,fang2025radioformer,deng2025mars,zhou2025tire}. Sparse Bayesian and group-sparse formulations are especially relevant to source identification under dictionary mismatch \cite{10264147,dall2012group}, and local-RM source localization addresses a related multi-emitter output \cite{11261889}. Recent satellite-specific work such as DeepRM reconstructs multidimensional LEO radio maps from limited samples through neural tensor decomposition \cite{yuan2025deeprm}; Generative Spectrum Cartography considers reconstruction and active sensing, including satellite-monitoring settings \cite{gu2026gsc}. {\color{black}These methods supply aggregate-map comparators, whereas SateBeam jointly returns the active satellite set and its reconstructed map. Identity metrics therefore evaluate support-producing methods, and query-map metrics evaluate every comparator.}

Sparse pursuit is a natural identity baseline when each candidate has a dictionary of possible beams. Its results depend strongly on grid resolution and dictionary memory. Continuous parametric fitting avoids fixed-grid bias but introduces nonconvex initialization and order-selection issues. SateBeam combines the two roles explicitly: the grid proposes a candidate, continuous refinement decides whether that candidate improves a complete physical model, and the same Gamma-BIC accounts for every accepted source.
\section{System Model and Problem Formulation}
\label{sec:system}

\subsection{Observer Information Model}
We consider a ground monitoring system that measures co-channel downlink power over a target region during $T$ short observation epochs. {\color{black}Public two-line elements (TLEs), propagated satellite positions and velocities, receiver coordinates, timestamps, carrier frequency, measurement bandwidth, and nominal receiver calibration are available. The candidate ephemerides are therefore known.} The contemporaneous active subset, its cardinality, operational beam directions, beamwidths, and effective in-band radiated powers remain unknown. SateBeam estimates these quantities from ground power measurements conditional on the known ephemerides.

{\color{black}The input consists of averaged received power. Beamforming and precoding enter the observation through the resulting radiated power pattern.} Signals from distinct satellites are modeled as uncorrelated over the energy-averaging interval, so their average powers add. This downlink radiation-interface model covers transparent and regenerative payloads through their effective in-band patterns over the observation interval.

\begin{figure*}[t]
\centering
\includegraphics[width=\textwidth]{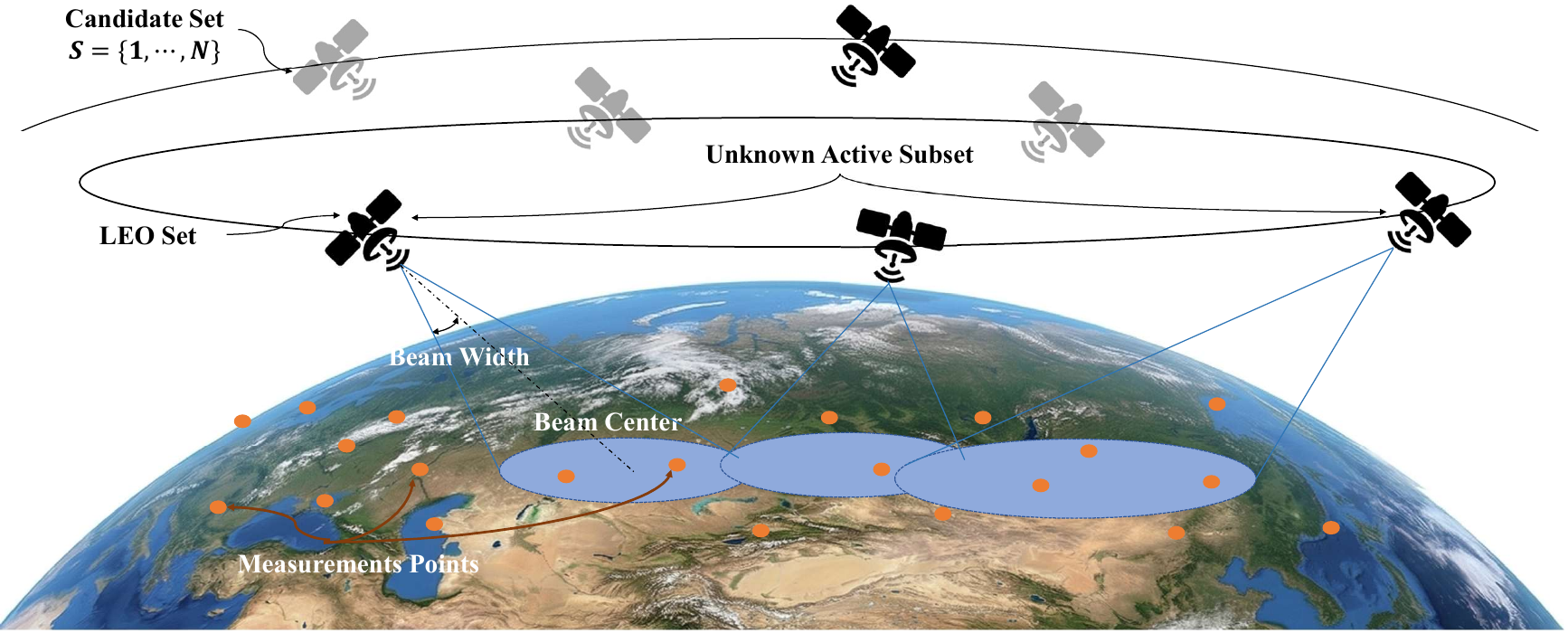}
\caption{System overview with known ephemeris candidates, unknown active beams, and sparse ground measurements.}
\label{fig:system_overview}
\end{figure*}

\subsection{Geometry, Visibility, and Direction Coordinates}
Let $t\in\mathcal T=\{1,\ldots,T\}$ index epochs, $m\in\mathcal M=\{1,\ldots,M\}$ index ground locations, and $s\in\mathcal S=\{1,\ldots,N_0\}$ index ephemeris candidates. In an Earth-centered Earth-fixed frame, the satellite and receiver positions are $\mathbf p_s(t)$ and $\mathbf r_m$. Define
\begin{align}
d_{tms}&=\|\mathbf r_m-\mathbf p_s(t)\|_2, &
\mathbf u_{tms}&=\frac{\mathbf r_m-\mathbf p_s(t)}{d_{tms}},
\label{eq:geometry}
\end{align}
where $\mathbf u_{tms}$ points from the satellite to the receiver. Let $\mathbf n_m$ denote the local outward ground normal and $\mathbf z_s(t)=-\mathbf p_s(t)/\|\mathbf p_s(t)\|_2$ the satellite nadir direction. The ground elevation $e_{tms}$ and satellite off-nadir angle $\psi_{tms}$ are
\begin{align}
e_{tms}&=\arcsin\!\left[(-\mathbf u_{tms})^{\mathsf T}\mathbf n_m\right], &
\psi_{tms}&=\arccos\!\left[\mathbf u_{tms}^{\mathsf T}\mathbf z_s(t)\right].
\label{eq:angles}
\end{align}
The inner product is the Euclidean dot product between vectors expressed in the same frame; inverse-trigonometric arguments are clipped to $[-1,1]$ numerically. Thus, elevation, off-nadir angle, and separation from the fitted beam center have distinct definitions.

A candidate is retained when it satisfies the minimum elevation $e_{\min}$, the steering limit $\psi_{\max}$, and the minimum spatial-support fraction $\rho_{\min}$:
\begin{equation}
\begin{split}
\widetilde{\mathcal S}=\biggl\{s\in\mathcal S:\;&
\frac{1}{TM}\sum_{t,m}\ind\{e_{tms}\ge e_{\min},\\
&\psi_{tms}\le\psi_{\max}\}\ge\rho_{\min}\biggr\}.
\end{split}
\label{eq:visibility}
\end{equation}
We write $N=|\widetilde{\mathcal S}|$ and denote the unknown active set by $\mathcal A\subseteq\widetilde{\mathcal S}$, with unknown cardinality $K=|\mathcal A|$. {\color{black}The support is an unordered set: the mean model sums the candidate contributions, so relabeling the elements of $\mathcal A$ leaves the likelihood, reconstructed map, and reported support unchanged.} This screen is determined exclusively by geometry.

For beam fitting, each satellite direction is represented by the unit vector $\mathbf q_{tms}$ in a satellite-centered frame. The reference direction $\mathbf c_{s,0}(t)$ points toward the ground-region center. Two bounded steering offsets $(\delta_{1s},\delta_{2s})$ define a fitted unit center $\mathbf c_s(t)$ through a local tangent chart. With orthonormal tangent directions $\mathbf a_s(t)$ and $\mathbf b_s(t)$,
\begin{align}
\mathbf v_s(t)&=\mathbf c_{s,0}(t)+\tan\delta_{1s}\mathbf a_s(t)
+\tan\delta_{2s}\mathbf b_s(t),\notag\\
\mathbf c_s(t)&=\mathbf v_s(t)/\|\mathbf v_s(t)\|_2.
\label{eq:center_chart}
\end{align}
The exact angular separation used by the circular-aperture model is
\begin{equation}
\theta_{tms}=\arccos\!\left[\mathbf q_{tms}^{\mathsf T}\mathbf c_s(t)\right].
\label{eq:angular_sep}
\end{equation}
This construction explains the coordinate transform before it is used: the offsets parameterize steering in two local cuts, while $\theta_{tms}$ is invariant to the particular chart and is used for beam gain and center-error reporting.

\begin{figure}[t]
\centering
\begin{tikzpicture}[x=1.20cm,y=0.88cm,>=stealth, every node/.style={font=\footnotesize}]
\draw[thick] (-2.0,0) .. controls (0,0.42) .. (2.4,0);
\node at (0,-0.32) {ground region};
\coordinate (S) at (1.6,3.5);
\coordinate (G) at (-1.5,0.13);
\coordinate (C) at (0.3,0.21);
\fill (S) circle (2pt) node[above] {$\mathbf p_s(t)$};
\fill (G) circle (2pt) node[left] {$\mathbf r_m$};
\fill (C) circle (1.6pt);
\draw[->,thick] (S)--(G) node[midway,left] {$d_{tms},\mathbf u_{tms}$};
\draw[->,dashed] (S)--(1.6,1.1) node[right] {nadir $\mathbf z_s$};
\draw[->] (S)--(C) node[pos=0.70,right] {$\mathbf c_s$};
\draw[dotted] (G)--(-0.1,0.13);
\draw[->,dashed] (G)--(-1.5,1.0) node[left] {$\mathbf n_m$};
\draw (-1.07,0.13) arc (0:47:0.43);
\node at (-0.85,0.38) {$e_{tms}$};
\draw (1.6,2.8) arc (-90:-133:0.70);
\node at (1.02,2.73) {$\psi_{tms}$};
\draw[dashed] (0.3,0.20) ellipse (0.62 and 0.15);
\node[align=left] at (1.25,0.55) {beam\ footprint};
\end{tikzpicture}
\caption{SateBeam geometry and beam-direction quantities.}
\label{fig:geometry}
\end{figure}

\subsection{Beam Pattern and Explicit Link Budget}
{\color{black}The nominal estimator uses the normalized uniform circular-aperture pattern from classical aperture theory and adopted for NTN studies in 3GPP TR~38.811 \cite{balanis2016antenna,tr38811}. We parameterize it directly by full half-power beamwidth (HPBW) $\beta_s$:}

\begin{align}
g_{tms}(\theta_{tms};\beta_s)&=
\begin{cases}
1, & z_{tms}=0,\\
\left[2J_1(z_{tms})/z_{tms}\right]^2,&z_{tms}\ne0,
\end{cases}\label{eq:circular_gain}\\
z_{tms}&=x_{1/2}\frac{\sin\theta_{tms}}{\sin(\beta_s/2)},\notag\\
\left[2J_1(x_{1/2})/x_{1/2}\right]^2&=\frac12,
\label{eq:circular_hpbw}
\end{align}

where $J_1$ is the first-order Bessel function and $x_{1/2}$ is the unique positive half-power root before the first null. Hence $g(0)=1$ and $g(\beta_s/2)=1/2$, making $\beta_s$ a physically interpretable HPBW. The mismatch experiments use separable squared-sinc, rotated elliptical, shadowed, and blocked truth fields with \eqref{eq:circular_gain} retained in the estimator.

{\color{black}Let $E_s$ be the peak-direction in-band EIRP in watts. The spatially varying nominal link factor is

\begin{align}
\ell_{tms}&=G_{r,m}(\mathbf u_{tms},f_c)
\left(\frac{c_0}{4\pi f_cd_{tms}}\right)^2\notag\\
&\quad\times10^{-\left(L_{\mathrm{atm},tms}+L_{\mathrm{pol},tms}\right)/10},
\label{eq:link_factor}\\
p_{tms}&=\alpha_s\ell_{tms}g_{tms}(\theta_{tms};\beta_s),
\label{eq:link_budget}
\end{align}}

where $G_{r,m}$ is receiver gain, $L_{\mathrm{atm},tms}$ and $L_{\mathrm{pol},tms}$ are atmospheric and polarization losses in dB, and $c_0$ is the speed of light. Free-space loss, frequency, slant range, receiver gain, atmospheric loss, and polarization loss are explicit. The dimensionless $\alpha_s$ captures a satellite-specific constant EIRP offset within the stated $\pm3$ dB bound. Residual spatial calibration, shadowing, and blockage are represented as receiver-dependent perturbations, with independently generated shadowing and blockage fields used in the robustness evaluation. ITU-R P.676/P.618 provide the corresponding detailed gas and Earth--space attenuation formulations \cite{itu676,itu618}.

\subsection{Doppler-Resolved Averaged-Power Observation}
Known ephemerides also provide radial velocity $\dot d_{tms}$ and the predicted Doppler shift

\begin{equation}
f_{D,tms}=-\frac{f_c}{c_0}\dot d_{tms}.
\label{eq:doppler}
\end{equation}

The monitored bandwidth $B$ is divided into $F$ subbands with edges $(f_{b-1},f_b)$. For a unit-integral Gaussian source spectrum $S_s(f)$, the captured fraction is
\begin{equation}
\chi_{tmsb}=\int_{f_{b-1}}^{f_b}S_s(f-f_{D,tms})\,df.
\label{eq:spectral_capture}
\end{equation}
This term makes Doppler useful for source separation and accounts for finite-band capture. A pure frequency shift does not change total received power when the full spectrum is captured; it changes the distribution across finite subbands.

{\color{black}Let $n_b$ be calibrated mean thermal-noise power in subband $b$, and let $b_0\ge0$ be a constant additional in-band background estimated from the observations. The conditional mean energy is

\begin{equation}
\mu_{tmb}=n_b+b_0+\sum_{s\in\mathcal A}
\alpha_s\ell_{tms}g_{tms}(\theta_{tms};\beta_s)\chi_{tmsb}.
\label{eq:mean_power}
\end{equation}}

{\color{black}The same $b_0$ is fitted in the null model, every candidate model, and all joint refinements. The estimator uses one shared background parameter, distinct from the calibrated thermal-noise means $n_b$, to represent spatially uniform additional power.}

{\color{black}Each reported measurement averages $L$ independent complex-voltage energy samples. Under the working model,

\begin{align}
y_{tmb}\mid\mu_{tmb}&\sim\operatorname{Gamma}(L,\mu_{tmb}/L),\notag\\
\mathbb E[y_{tmb}]&=\mu_{tmb},\qquad
\operatorname{var}(y_{tmb})=\mu_{tmb}^2/L.
\label{eq:gamma_measurement}
\end{align}}

This explicitly states the statistical model for averaged power. The model assigns receiver-dependent propagation and calibration perturbations separately from calibrated thermal noise, and the robustness study independently varies the named spatial propagation fields.

\subsection{Inference Target and Working Order Criterion}
Let $n=TMF$ and stack all observations in $\mathbf y\in\mathbb R_+^n$. For support $\mathcal A$ and parameters $\Theta=\{\alpha_s,\delta_{1s},\delta_{2s},\beta_s\}_{s\in\mathcal A}$, denote the stacked mean by $\boldsymbol\mu(\mathcal A,\Theta,b_0)$. The Gamma deviance up to constants independent of the model is

\begin{equation}
\mathcal D(\mathcal A,\Theta,b_0)=2L\sum_{i=1}^{n}
\left(\log\mu_i+\frac{y_i}{\mu_i}\right).
\label{eq:gamma_deviance}
\end{equation}

{\color{black}The working model-order objective is

\begin{problem}\label{prob:joint}
\begin{align}
\min_{\mathcal A,\Theta,b_0}\quad &
\mathcal D(\mathcal A,\Theta,b_0)+(4|\mathcal A|+1)\log n\notag\\
\mathrm{s.t.}\quad &\mathcal A\subseteq\widetilde{\mathcal S},\quad
|\mathcal A|\le K_{\max},\quad b_0\ge0,\notag\\
&10^{-0.3}\le\alpha_s\le10^{0.3},\quad
|\delta_{1s}|,|\delta_{2s}|\le6^\circ,\notag\\
&4^\circ\le\beta_s\le12^\circ.
\end{align}
\end{problem}}

{\color{black}Each accepted source contributes three beam parameters and one amplitude. The BIC count includes the profiled amplitude as an estimated parameter. The common background appears in both nested models and cancels in an incremental comparison.}

At a query location $\mathbf r$, SateBeam evaluates the same forward model with the propagated geometry to obtain a satellite-induced map and a total received-power map. Query points are reserved for evaluation. Separate identification and map metrics assess recovered satellite identities and spatial prediction, respectively.

{\color{black}Gamma-BIC defines the model-order rule for the adaptive search. Candidate selection and the zero-amplitude boundary determine its statistical setting. Null false-alarm and rule-sensitivity experiments directly validate the declared $4\log n$ penalty.}

\section{SateBeam Solver}
\label{sec:method}

The inverse problem couples a large discrete ephemeris set with a small continuous beam model. The full spatial dictionary grows with the visible set, and grid-first support selection introduces off-grid bias. {\color{black}SateBeam addresses both effects with a four-stage solver: ephemeris-derived Doppler screening, GPU grid proposals, continuous post-proposal refinement with Gamma-BIC order selection, and forward--backward support correction. Screening controls dictionary size, post-proposal refinement preserves weak off-grid sources, backward deletion removes redundant components, and support exchange revisits early greedy choices.}

\subsection{All-Candidate Doppler NNLS Screening}
For each space--time sample $(t,m)$, collect the $F$ measured subband energies in $\mathbf y_{tm}\in\mathbb R_+^F$. Known ephemerides and \eqref{eq:doppler} determine the normalized spectral vector $\boldsymbol\chi_{tms}\in\mathbb R_+^F$ for every candidate. SateBeam solves
\begin{equation}
\min_{\mathbf z_{tm}\ge0,\,c_{tm}\ge0}
\left\|\mathbf y_{tm}-\sum_{s=1}^{N}z_{tm,s}\boldsymbol\chi_{tms}
-c_{tm}\mathbf1\right\|_2^2
\label{eq:spectral_nnls}
\end{equation}
and assigns candidate $s$ the root-mean-square score
\begin{equation}
r_s=\left(\frac{1}{TM}\sum_{t,m}z_{tm,s}^2\right)^{1/2}.
\label{eq:screen_score}
\end{equation}
No beam center, beamwidth, activity label, or query sample is used. The retained pool consists of the $P$ largest scores, where
\begin{align}
g_N&=\max\!\left\{2,\left\lceil\log_2
\frac{N}{K_{\mathrm{cap}}}\right\rceil\right\},\notag\\
P&=\min\!\left\{N,K_{\mathrm{cap}}+g_N\right\}.
\label{eq:screen_pool}
\end{align}
{\color{black}The logarithmic guard retains additional near-ties as the candidate set grows. With $K_{\mathrm{cap}}=8$, the nominal $N=16$ case retains ten candidates and $N=64$ retains eleven.}

\subsection{GPU Grid Proposal and Profiled Coefficients}
For each retained satellite, steering offsets are proposed on a $2^\circ$ grid over $[-6^\circ,6^\circ]^2$, and HPBW is proposed at five equally spaced values in $[4^\circ,12^\circ]$. Thus, each candidate contributes $7\times7\times5=245$ atoms. All atoms use the same explicit link budget, circular-aperture pattern, Doppler weights, background treatment, and observation samples as the continuous estimator.

{\color{black}At greedy iteration $k$, let $\mathcal A_k$ be the current support and $\mathbf a_j$ a normalized grid atom from an unselected candidate. The GPU computes all residual correlations in one matrix operation and proposes the largest positive correlation. The grid supplies a candidate and a starting point for continuous beam refinement.}

For fixed beam parameters, source amplitudes and background are obtained from bounded least squares:
\begin{equation}
(\widehat{\boldsymbol\alpha},\widehat b_0)=
\arg\min_{\boldsymbol\alpha,b_0}
\left\|\mathbf y-\mathbf n-b_0\mathbf1-
\sum_{s\in\mathcal A_k}\alpha_s\mathbf a_s\right\|_2^2,
\label{eq:profile_ls}
\end{equation}
subject to the bounds in Problem~\ref{prob:joint}. A newly proposed component is allowed to take zero amplitude before acceptance. {\color{black}Every solve includes the same constant column, ensuring identical background treatment in the null and candidate models.}

\subsection{Continuous Refinement Before Gamma-BIC Selection}
{\color{black}Continuous refinement before order scoring removes grid-placement bias for weak sources whose centers lie between proposal points. SateBeam appends the proposal and jointly refines all selected steering offsets, HPBWs, amplitudes, and background with bounded L-BFGS-B \cite{nocedal2006numerical}. A finite trial update that improves the profiled fitting objective advances to order scoring.}

Let $\mathcal D_k$ and $\mathcal D_{k+1}(s)$ denote the Gamma deviances in \eqref{eq:gamma_deviance} after continuous refinement. Because one source contributes four parameters, the incremental score is
\begin{equation}
\Delta_{\mathrm{BIC}}(s)=\mathcal D_k-\mathcal D_{k+1}(s)-4\log n.
\label{eq:bic_score}
\end{equation}
The best proposal is accepted for $\Delta_{\mathrm{BIC}}>0$, and forward selection terminates at the first nonpositive score. Standard BIC supplies the main results. AIC replaces $4\log n$ by $8$, and margin-BIC requires $\Delta_{\mathrm{BIC}}>\tau$ in the labeled sensitivity study. The main configuration is fully specified by the Gamma deviance, physical parameter bounds, and zero additional selection margin.

\subsection{Backward Deletion and Support Exchange}
{\color{black}Forward selection is path dependent. After it terminates, SateBeam evaluates deletion of each selected component with the remaining beams fixed, chooses the most promising deletion, and fully refines that reduced model. A lower complete Gamma-BIC accepts the deletion, and the procedure repeats to a deletion-stable support.}

{\color{black}Support exchange revisits early greedy choices through residual-guided one-for-one replacements. SateBeam removes each selected source in turn, proposes the strongest unselected residual atom, and continuously refines the replacement. The lowest-Gamma-BIC improvement is accepted using the declared observations and ephemeris candidates.}

\begin{algorithm}[t]
\caption{SateBeam with Doppler screening and support correction}
\label{alg:satebeam}
\begin{algorithmic}[1]
\Require averaged subband powers $\mathbf y$, known ephemerides, link-budget configuration, $K_{\mathrm{cap}}$
\Ensure support $\widehat{\mathcal A}$, beam parameters, background, and query map
\State Apply the geometry screen \eqref{eq:visibility}.
\State {\color{black}Solve \eqref{eq:spectral_nnls}; retain the pool \eqref{eq:screen_pool}.}
\State Build the bounded beam dictionary for the retained candidates.
\State Fit the background-only model; set $\mathcal A_0\gets\varnothing$.
\For{$k=0,\ldots,K_{\max}-1$}
  \State Propose the unselected satellite/grid atom with maximum positive residual correlation.
  \State {\color{black}Jointly refine the trial support and compute \eqref{eq:bic_score}.}
  \If{$\Delta_{\mathrm{BIC}}\le0$}
    \State \textbf{break}
  \EndIf
  \State Accept the source and update the residual.
\EndFor
\Repeat
  \State {\color{black}Screen all one-source deletions; fully refine the best reduced model.}
\Until{no deletion lowers Gamma-BIC}
\Repeat
  \State {\color{black}Test the best residual-guided one-for-one support replacement.}
\Until{no replacement lowers Gamma-BIC}
\State Perform final bounded joint refinement and synthesize the independent-query map.
\end{algorithmic}
\end{algorithm}

\subsection{Identifiability}
{\color{black}Source identifiability depends on the separation of feasible ephemeris-conditioned templates. We remove the fitted constant-background direction from two sampled templates and normalize them. Their absolute inner product defines background-projected coherence.} Active--active coherence measures separation within the true support, and active--inactive coherence measures separation from competing candidates. We report exact support, F1, angular separation, both coherence measures, and query RMSE together across equal and unequal source strengths.

\subsection{Computational Cost}
{\color{black}Let $n=TMF$, $G=G_c^2G_\beta$ be the atoms per retained satellite, $D=PG$ the dictionary size, and $\widehat K\le K_{\max}$ the selected order. Doppler screening costs $TM\,C_{\mathrm{NNLS}}(F,N+1)$ for the chosen nonnegative least-squares implementation. Dictionary construction costs $O(nD)$ time and memory. Across forward iterations, GPU correlation costs $O(\widehat K nD)$ arithmetic, while small bounded coefficient solves add $\sum_k C_a(n,k+2)$.}

For a $k$-source joint model, one objective evaluation costs $O(nk)$. If $I_k$ evaluations are used, forward refinement costs
\begin{equation}
O\!\left(\sum_{k=1}^{\widehat K+1} I_k n k\right).
\label{eq:complexity_joint}
\end{equation}
Backward screening uses $O(n\widehat K^2)$ fixed-parameter work plus at most one fully refined deletion per round. Support exchange uses $O(nD\widehat K)$ correlation work and refines only the best proposed replacements. These expressions distinguish the Doppler screen's dependence on $N$ from the spatial dictionary's dependence on the retained pool $P$.

In the nominal $N=16$ configuration, $n=4\times160\times33=21{,}120$, $P=10$, $G=245$, and $D=2{,}450$. The float64 physical dictionary occupies approximately 414 MB, and GPU normalization uses a single float32 representation. At $N=64$, SateBeam retains eleven candidates and completes within the 8 GB evaluation GPU; the corresponding complete OMP dictionary requires more than 8 GB. Empirical latency and dictionary construction are reported together with the accuracy curves.

\section{Experimental Evaluation}
\label{sec:experiments}

\subsection{Reproducible Setup and Protocol}
The experiment combines real orbital inputs with controlled received-power measurements. We use a fixed public Starlink TLE snapshot containing 10,713 records and propagate it with SGP4. The median catalog epoch is 2026-09-10 10:07:05 UTC. Each scene uses a deterministic epoch offset and draws a geometrically feasible candidate roster after applying the declared elevation, steering, altitude, and spatial-support constraints. Real ephemeris-derived positions and velocities define the orbital geometry; controlled simulation supplies activity, beam parameters, fading, and energy measurements for paired evaluation.

\begin{table}[t]
\centering
\scriptsize
\caption{Core experimental and solver settings.}
\label{tab:settings}
\setlength{\tabcolsep}{1.5pt}
\renewcommand{\arraystretch}{1.02}
\begin{tabular}{@{}>{\raggedright\arraybackslash}p{0.22\columnwidth}>{\raggedright\arraybackslash}p{0.43\columnwidth}>{\raggedright\arraybackslash}p{0.30\columnwidth}@{}}
\toprule
Item & Setting & Role \\
\midrule
Orbit knowledge & Frozen Starlink TLE; SGP4 ephemeris & Known ephemeris \\
Nominal dimensions & $N=16$, $K=4$, $M=160$ & Four snapshots \\
Snapshot spacing & 60 s & 240 s observation span \\
RF setup & 12 GHz, 1 MHz, 33 subbands & 50 kHz source FWHM \\
Nominal EIRP & 36 dBW & Fixed engineering test value \\
Energy averaging & 64 looks & Gamma likelihood \\
Geometry screen & $h=200$--$2000$ km; $e\ge20^\circ$, $\psi\le65^\circ$, $\rho\ge0.35$ & Visibility/support \\
Beam width & 4--12$^\circ$ & Truth and estimator \\
Center offset & $\pm3^\circ$ truth; $\pm6^\circ$ fit & Off-grid refinement \\
Power bounds & $\alpha_s$: $\pm3$ dB; $0\le b_0\le5n_b$ & Amplitude/background \\
Grid & 2$^\circ$; five width levels & Initialization only \\
Screening & 10 candidates nominal; 11 at $N=64$ & $K_{\max}=8$ \\
Optimizer & Bounded L-BFGS-B & Joint continuous refinement \\
Model order & Gamma-BIC; penalty $4\log n$; margin 0 & No robust loss/weights \\
Baselines & Shared observations, domains, background & Method-specific settings released \\
\bottomrule
\end{tabular}
\end{table}

The nominal case uses $N=16$ candidates, $K=4$ active satellites, $M=160$ measured locations, $T=4$ epochs spaced by 60 s, and $F=33$ subbands over 1 MHz. A $32\times32$ ground grid leaves 864 independent query locations per epoch, or 3,456 query predictions per scene. Query locations are excluded from fitting, screening, stopping, and parameter selection. The nominal EIRP is 36 dBW, receiver gain is 10 dBi, noise figure is 5 dB, zenith atmospheric loss is 0.2 dB, polarization loss is 0.5 dB, and $L=64$ energy looks. Truth and estimator use the circular-aperture model in the matched case; independent mismatch conditions alter only the named factor.

{\color{black}The primary comparison uses 30 paired scenes. Each method receives the same observations, geometry, train/query split, and random realization. We report scene-bootstrap 95\% confidence intervals and paired two-sided Wilcoxon tests.} Model-mismatch experiments use 20 paired scenes per condition, and $N$ and $K$ scale sweeps use ten. A separate ten-scene serial GPU run after warm-up supplies the reported SateBeam latency.

Each archived scene records the propagated states, train/query masks, physical truth, perturbations, and observations. Each method record stores the fitted support and parameters, predictions, selection trace, runtime, and solver status, while the released summaries preserve one row per completed scene in every reported condition.

\subsection{Physical and Dynamic Validation}
{\color{black}Before stochastic evaluation, the implementation is checked against analytical identities. Doubling range at fixed frequency and gain, and doubling frequency at fixed range and normalized gain, each changes free-space loss by 6.0206 dB. The hand-computed and implemented single-link budgets agree to numerical precision. Treating path loss as one satellite-wide constant creates mean spatial errors of 0.13, 0.52, and 1.31 dB over 50, 200, and 500 km regions, respectively, validating the explicit $d_{tms}$ term in \eqref{eq:link_factor}.}

\begin{table}[t]
\centering
\footnotesize
\caption{Physical and motion validation.}
\label{tab:physical_checks}
\setlength{\tabcolsep}{3.4pt}
\begin{tabular}{lcc}
\toprule
Check & Setting & Result \\
\midrule
Range doubling & 500 to 1000 km & 6.0206 dB \\
Frequency doubling & 10 to 20 GHz & 6.0206 dB \\
Constant-range error & 200 km region & 0.52 dB \\
Ground-fixed beam & 10 s window & 0.01 dB \\
Satellite-fixed beam & 10 s window & 0.39 dB \\
Doppler at 12 GHz & 50 scenes & 144 kHz mean \\
1 MHz capture loss & 50 scenes & 0.69 dB mean \\
\bottomrule
\end{tabular}
\end{table}

{\color{black}Fifty ephemeris scenes are also propagated through the observation interval. At 10 s, the midpoint-static approximation changes total power by 0.01 dB for a ground-fixed beam and 0.39 dB for a satellite-frame-fixed beam. Mean Doppler magnitude at 12 GHz is approximately 144 kHz, and a fixed 1 MHz rectangular receiver loses 0.69 dB of captured power on average.} These results motivate the four propagated epochs and explicit subband capture in the main protocol.

\subsection{Comparators, Metrics, and Dictionary Fairness}
The source-identification comparators are full-dictionary OMP, nonnegative Lasso, an orbit-centered nonnegative least-squares reference (Orbit-NNLS), and peak detection. Their atoms use the same link budget, circular-aperture pattern, steering domain, width domain, observations, and fitted background as SateBeam. OMP and Lasso use the same $2^\circ$ center grid and five widths. The oracle-support fit provides a reference with known identities. It receives the true support and estimates all continuous parameters and background from the same observations.

Map-only comparators are Kriging, SoftImpute, and a documented two-dimensional, single-frequency adaptation of the unsupervised neural tensor-decomposition stage of DeepRM \cite{yuan2025deeprm}. Their aggregate-map outputs are evaluated by query-map error and latency. Generative Spectrum Cartography and other recent satellite-monitoring map approaches motivate this comparison class \cite{gu2026gsc}. Support metrics evaluate methods that return active-satellite identities.

Detection metrics are precision, recall, F1, exact support, and absolute order error. Map quality is measured at independent query locations by total-power RMSE in dB and signal NMSE. Beam-center error is the great-circle separation between true and fitted center vectors; HPBW error is absolute width error. Parameter errors are computed on correctly identified satellites and reported together with recall. Runtime includes screening, dictionary generation, coefficient fitting, continuous refinement, and map synthesis.

\begin{figure}[t]
\centering
\includegraphics[width=\columnwidth]{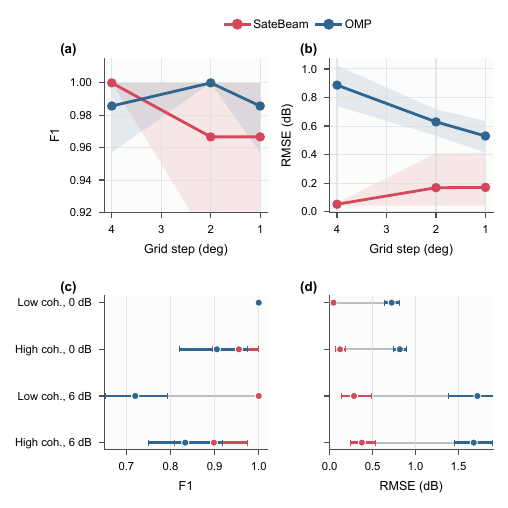}
\caption{Grid-resolution and empirical-coherence sensitivity; bars/bands show 95\% CIs.}
\label{fig:grid_identifiability}
\end{figure}

{\color{black}To isolate dictionary effects, center steps $4^\circ$, $2^\circ$, and $1^\circ$ are tested with the same offset and HPBW domains. At these three resolutions, SateBeam gives query RMSE $0.053/0.169/0.171$ dB, compared with OMP's $0.887/0.630/0.531$ dB. The corresponding reductions are 94.0\%, 73.2\%, and 67.8\%, establishing the map advantage at every tested grid resolution.} Figure~\ref{fig:grid_identifiability}(a,b) reports both F1 and RMSE for the complete paired sweep. The main $2^\circ$ comparison uses the same grid for both methods and delivers F1 $1.000$ with 100\% exact support for SateBeam.

\subsection{Main Paired Comparison}
\begin{table*}[t]
\centering
\definecolor{SateBeamTint}{HTML}{E7F0F7}
\definecolor{GainTint}{HTML}{EFEFEF}
\definecolor{SateBeamRed}{HTML}{A00000}
\caption{Main paired comparison over 30 real-orbit scenes; intervals are 95\% bootstrap CIs and runtime is median/P90.}
\label{tab:main_comparison}
\setlength{\tabcolsep}{1.2pt}
\renewcommand{\arraystretch}{1.08}
\begin{tabular}{lccccccc}
\toprule
Method & F1 $\uparrow$ [95\% CI] & Exact $\uparrow$ & RMSE $\downarrow$ [95\% CI] (dB) & Corr. $\uparrow$ & Center $\downarrow$ ($^\circ$) & HPBW $\downarrow$ ($^\circ$) & Time $\downarrow$ (s) \\
\midrule
\rowcolor{SateBeamTint} \textbf{SateBeam} & \textcolor{SateBeamRed}{\bfseries 1.000 [1.000, 1.000]} & \textcolor{SateBeamRed}{\bfseries 1.00} & \textcolor{SateBeamRed}{\bfseries 0.050 [0.044, 0.056]} & \textcolor{SateBeamRed}{\bfseries 0.9999} & \textcolor{SateBeamRed}{\bfseries 0.06} & \textcolor{SateBeamRed}{\bfseries 0.06} & 14.57 / 29.60 \\
OMP & 0.975 [0.951, 0.993] & 0.83 & 0.655 [0.580, 0.729] & 0.9775 & 1.07 & 0.94 & 8.58 / 9.63 \\
Lasso & 0.661 [0.650, 0.667] & 0.00 & 2.248 [2.015, 2.487] & 0.8627 & 1.91 & 2.75 & 9.42 / 10.76 \\
Orbit-NNLS & 0.897 [0.854, 0.937] & 0.47 & 1.709 [1.561, 1.847] & 0.8801 & 1.95 & 2.06 & 0.09 / 0.12 \\
Peak & 0.048 [0.000, 0.113] & 0.00 & 7.963 [7.270, 8.543] & 0.0806 & 1.88 & 2.72 & 0.04 / 0.06 \\
DeepRM-TD & -- & -- & 0.253 [0.228, 0.279] & 0.9968 & -- & -- & 3.54 / 4.53 \\
Kriging & -- & -- & 0.353 [0.326, 0.380] & 0.9941 & -- & -- & 0.03 / 0.04 \\
SoftImpute & -- & -- & 2.933 [2.775, 3.091] & 0.3483 & -- & -- & 0.03 / 0.04 \\
Oracle & 1.000 [1.000, 1.000] & 1.00 & 0.050 [0.044, 0.056] & 0.9999 & 0.06 & 0.06 & 10.80 / 12.09 \\
\midrule\rowcolor{GainTint} \textbf{Rel. gain vs. best prior} & \textcolor{SateBeamRed}{\bfseries +2.6\%} & \textcolor{SateBeamRed}{\bfseries +20.0\%} & \textcolor{SateBeamRed}{\bfseries +80.2\%} & \textcolor{SateBeamRed}{\bfseries +0.3\%} & \textcolor{SateBeamRed}{\bfseries +94.1\%} & \textcolor{SateBeamRed}{\bfseries +93.8\%} & -- \\
\bottomrule
\end{tabular}
\end{table*}

{\color{black}SateBeam recovers the exact four-satellite support in all 30 scenes, giving F1 $1.000$ and exact-support rate 100\%. OMP gives F1 $0.975$ and exact support 83.3\%. The paired F1 difference is $+0.0253$ (two-sided Wilcoxon $p=0.0412$). SateBeam reduces query RMSE from OMP's 0.655 dB to 0.050 dB, a 92.4\% reduction relative to OMP ($p=1.86\times10^{-9}$), and matches the 0.050 dB oracle-support reference.} Its query correlation is 0.9999, and its mean beam-center and HPBW errors are $0.063^\circ$ and $0.059^\circ$. Relative to the strongest non-oracle method for each quality metric, the gains are 2.6\% in F1, 20.0\% in exact support, 80.2\% in query RMSE, 0.3\% in correlation, 94.1\% in center error, and 93.8\% in HPBW error.

\begin{figure}[t]
\centering
\includegraphics[width=\columnwidth]{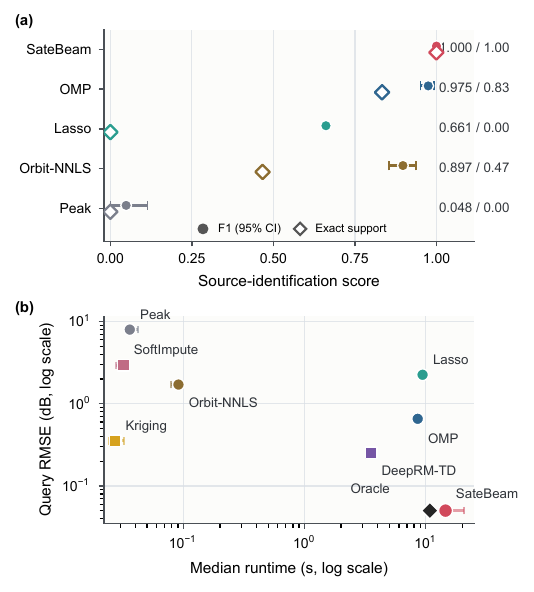}
\caption{Main comparison: (a) F1 and exact support; (b) runtime versus query RMSE. Bars show 95\% CIs.}
\label{fig:main_results}
\end{figure}

{\color{black}The solver ablation retains the same forward model, proposal grid, and Gamma-BIC rule and removes the scale-aware screening guard, continuous refinement before BIC, backward deletion, and support exchange. The serial GPU benchmark records SateBeam median/P90 latency of 14.57/29.60 s, improving the ablation's 15.60/34.18 s while converting all eight support errors to exact recovery. These measurements include the complete inference pipeline. SateBeam is the only non-oracle method that simultaneously reaches F1 $1.000$, 100\% exact support, and 0.050 dB query RMSE.}

The paired support and map tests provide complementary evidence for the joint inference objective. The F1 test measures satellite-identification accuracy, while the query-RMSE test measures spatial prediction at locations excluded from fitting. Both comparisons favor SateBeam under the declared paired protocol. The oracle-support reference further isolates the cost of inferring the active set. It receives the true identities and fits the continuous parameters on the same observations. SateBeam reaches the same reported 0.050 dB mean query RMSE while recovering the identities from the measurements. Together, exact-support recovery and independent-query accuracy establish the two outputs required for satellite-attributed radio mapping.

\begin{figure*}[!b]
\centering
\includegraphics[width=0.675\textwidth]{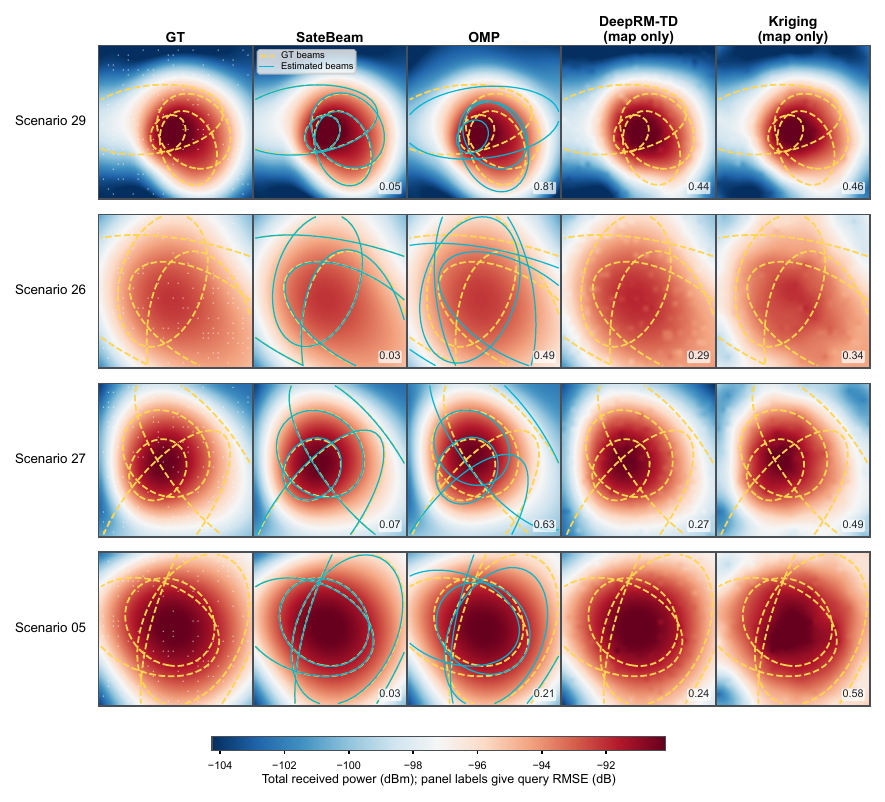}
\caption{Four RM comparisons with large query-error separation. Yellow dashed/cyan solid contours denote GT/estimated beams; labels give query RMSE (dB).}
\label{fig:rm_matrix}
\end{figure*}

\subsection{Qualitative Recovery Mechanisms}
\begin{figure*}[!b]
\centering
\includegraphics[width=0.675\textwidth]{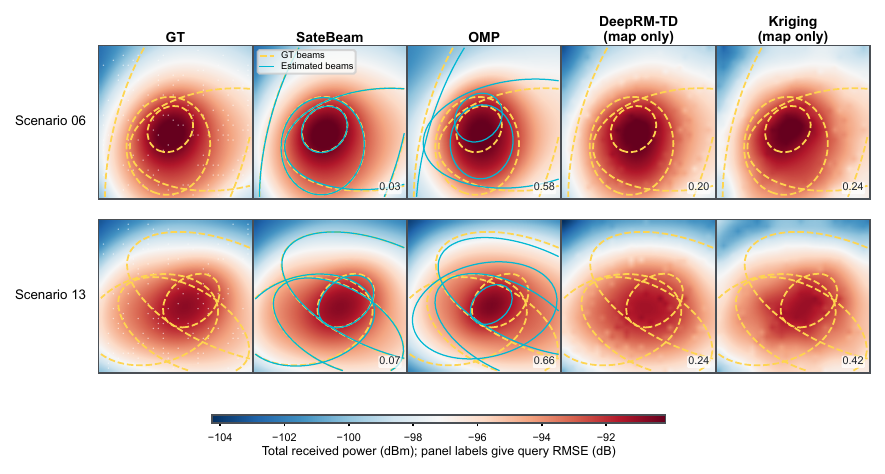}
\caption{Two additional RM comparisons with large query-error separation; layout follows Fig.~\ref{fig:rm_matrix}.}
\label{fig:rm_hardcases}
\end{figure*}

{\color{black}Figures~\ref{fig:rm_matrix} and~\ref{fig:rm_hardcases} show the six scenarios with the largest minimum query-RMSE separation between SateBeam and the three displayed baselines at the selected epoch. Across these scenarios, SateBeam attains 0.03--0.07 dB, compared with 0.21--0.81 dB for OMP, 0.20--0.44 dB for DeepRM-TD, and 0.24--0.58 dB for Kriging.} The absolute-power maps use one shared dBm scale, and every panel overlays the unordered GT half-power contours as yellow dashed curves; source-resolved estimates add cyan solid curves. The near-overlap in the SateBeam column and visible displacement in the OMP column expose the geometric difference directly. A common contour style preserves permutation invariance across satellite components.

\begin{figure*}[!b]
\centering
\includegraphics[width=0.70\textwidth]{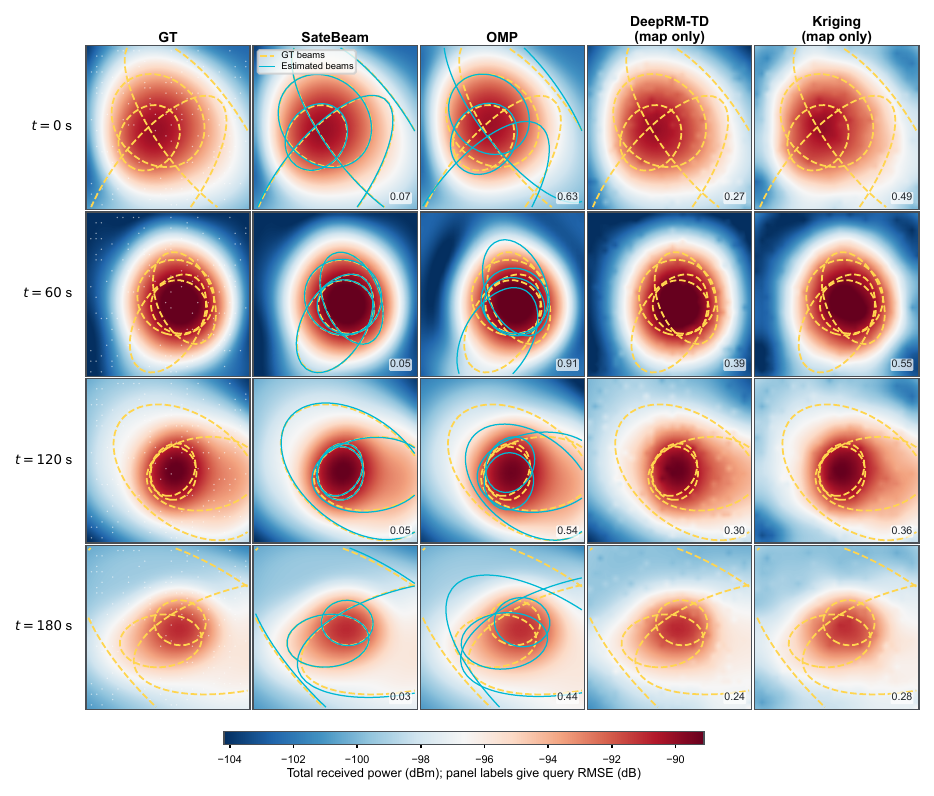}
\caption{Four-epoch RM reconstruction for Scenario 27. Columns follow Fig.~\ref{fig:rm_matrix}; labels give query RMSE (dB).}
\label{fig:rm_temporal}
\end{figure*}

{\color{black}Scenario 27 provides sustained separation across all four propagated epochs in Fig.~\ref{fig:rm_temporal}. SateBeam attains 0.03--0.07 dB, compared with 0.24--0.39 dB for the strongest displayed baseline at each epoch.} The paired GT/estimated contours track agreement as beam locations, shapes, and overlap evolve over 180 s, with permutation-invariant component display.

\subsection{Beam and Propagation Mismatch}
\begin{table}[t]
\centering
\definecolor{SateBeamTint}{HTML}{E7F0F7}
\definecolor{GainTint}{HTML}{EFEFEF}
\definecolor{SateBeamRed}{HTML}{A00000}
\scriptsize
\caption{Single-factor robustness over 20 paired scenes; gains are relative to OMP.}
\label{tab:mismatch}
\setlength{\tabcolsep}{1.2pt}
\renewcommand{\arraystretch}{1.06}
\resizebox{\columnwidth}{!}{%
\begin{tabular}{lccccccc}
\toprule
& \multicolumn{3}{c}{F1 $\uparrow$} & Exact $\uparrow$ & \multicolumn{3}{c}{RMSE $\downarrow$ (dB)} \\
\cmidrule(lr){2-4}\cmidrule(lr){6-8}
Condition & Sate & OMP & Gain & Sate/OMP (gain) & Sate & OMP & Red. \\
\midrule
Sinc pattern & \cellcolor{SateBeamTint}\textcolor{SateBeamRed}{\bfseries 1.000} & 0.975 & \cellcolor{GainTint}\textcolor{SateBeamRed}{\bfseries +2.6\%} & \shortstack{1.00/0.85\\\textcolor{SateBeamRed}{\bfseries +15.0 pp}} & \cellcolor{SateBeamTint}\textcolor{SateBeamRed}{\bfseries 0.20} & 0.71 & \cellcolor{GainTint}\textcolor{SateBeamRed}{\bfseries +71.4\%} \\
Elliptic beam & \cellcolor{SateBeamTint}\textcolor{SateBeamRed}{\bfseries 0.929} & 0.923 & \cellcolor{GainTint}\textcolor{SateBeamRed}{\bfseries +0.6\%} & \shortstack{0.50/0.50\\+0.0 pp} & \cellcolor{SateBeamTint}\textcolor{SateBeamRed}{\bfseries 1.38} & 1.53 & \cellcolor{GainTint}\textcolor{SateBeamRed}{\bfseries +9.7\%} \\
Shadowing (4 dB) & \cellcolor{SateBeamTint}\textcolor{SateBeamRed}{\bfseries 0.861} & 0.780 & \cellcolor{GainTint}\textcolor{SateBeamRed}{\bfseries +10.5\%} & \shortstack{0.40/0.05\\\textcolor{SateBeamRed}{\bfseries +35.0 pp}} & \cellcolor{SateBeamTint}\textcolor{SateBeamRed}{\bfseries 2.42} & 2.48 & \cellcolor{GainTint}\textcolor{SateBeamRed}{\bfseries +2.4\%} \\
Blockage (10\%) & \cellcolor{SateBeamTint}\textcolor{SateBeamRed}{\bfseries 0.944} & 0.837 & \cellcolor{GainTint}\textcolor{SateBeamRed}{\bfseries +12.8\%} & \shortstack{0.65/0.40\\\textcolor{SateBeamRed}{\bfseries +25.0 pp}} & \cellcolor{SateBeamTint}\textcolor{SateBeamRed}{\bfseries 3.22} & 3.29 & \cellcolor{GainTint}\textcolor{SateBeamRed}{\bfseries +2.1\%} \\
\bottomrule
\end{tabular}}
\end{table}

{\color{black}The estimator uses the same circular-aperture model in every condition. Sinc truth changes sidelobe structure, the rotated ellipse changes footprint shape, and correlated 4 dB shadowing and 10\% blockage introduce receiver-dependent spatial loss. SateBeam improves both F1 and query RMSE over OMP in all four cases. It gains 12.8\% in F1 under blockage and 10.5\% under shadowing, retains perfect support under sinc mismatch, and reduces query RMSE in every condition.} These paired gains establish robust source attribution and map reconstruction across independent beam and propagation perturbations.

\begin{figure}[!t]
\centering
\includegraphics[width=0.94\columnwidth]{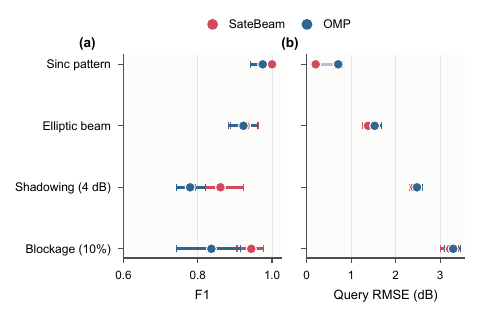}
\caption{Single-factor mismatch over 20 paired scenes; bars show 95\% CIs.}
\label{fig:mismatch}
\end{figure}

\subsection{Candidate and Active-Source Scaling}
\begin{table}[t]
\centering
\definecolor{SateBeamTint}{HTML}{E7F0F7}
\definecolor{GainTint}{HTML}{EFEFEF}
\definecolor{SateBeamRed}{HTML}{A00000}
\scriptsize
\caption{Scaling and gains over the solver ablation.}
\label{tab:scaling}
\setlength{\tabcolsep}{1.2pt}
\renewcommand{\arraystretch}{1.08}
\resizebox{\columnwidth}{!}{%
\begin{tabular}{lrrrrrrr}
\toprule
Sweep & Value & F1 $\uparrow$ & Exact $\uparrow$ & RMSE $\downarrow$ & F1 gain & Exact gain & RMSE red. \\
\midrule
Candidates $N$ & 8 & 1.000 & 1.00 & 0.03 & \cellcolor{GainTint}+0.0\% & \cellcolor{GainTint}+0.0 pp & \cellcolor{GainTint}+0.0\% \\
 & 16 & 0.986 & 0.90 & 0.06 & \cellcolor{GainTint}+0.0\% & \cellcolor{GainTint}+0.0 pp & \cellcolor{GainTint}+0.0\% \\
 & 32 & 1.000 & 1.00 & 0.03 & \cellcolor{GainTint}+0.0\% & \cellcolor{GainTint}+0.0 pp & \cellcolor{GainTint}+0.0\% \\
 & 64 & 0.953 & 0.70 & 0.14 & \cellcolor{GainTint}\textcolor{SateBeamRed}{\bfseries +16.7\%} & \cellcolor{GainTint}\textcolor{SateBeamRed}{\bfseries +40.0 pp} & \cellcolor{GainTint}\textcolor{SateBeamRed}{\bfseries +52.0\%} \\\addlinespace[2pt]
Active sources $K$ & 1 & 1.000 & 1.00 & 0.05 & \cellcolor{GainTint}+0.0\% & \cellcolor{GainTint}+0.0 pp & \cellcolor{GainTint}+0.0\% \\
 & 2 & 1.000 & 1.00 & 0.07 & \cellcolor{GainTint}+0.0\% & \cellcolor{GainTint}+0.0 pp & \cellcolor{GainTint}+0.0\% \\
 & 4 & 1.000 & 1.00 & 0.05 & \cellcolor{GainTint}\textcolor{SateBeamRed}{\bfseries +4.5\%} & \cellcolor{GainTint}\textcolor{SateBeamRed}{\bfseries +30.0 pp} & \cellcolor{GainTint}\textcolor{SateBeamRed}{\bfseries +70.8\%} \\
 & 8 & 0.944 & 0.80 & 0.22 & \cellcolor{GainTint}\textcolor{SateBeamRed}{\bfseries +1.6\%} & \cellcolor{GainTint}\textcolor{SateBeamRed}{\bfseries +30.0 pp} & \cellcolor{GainTint}\textcolor{SateBeamRed}{\bfseries +18.5\%} \\\addlinespace[2pt]
\bottomrule
\end{tabular}}
\vspace{1pt}\parbox{\columnwidth}{\scriptsize \emph{Ablation:} scale-aware screening guard, pre-BIC refinement, backward deletion, and support exchange removed.}
\end{table}

At $N=8,16,32,64$, SateBeam gives F1 $1.000,0.986,1.000,0.953$ and query RMSE $0.031,0.064,0.028,0.141$ dB. {\color{black}At $N=64$, it improves F1 by 16.7\%, exact support by 40 percentage points, and RMSE by 52.0\% relative to the defined solver ablation while completing with an eleven-candidate retained pool on the 8 GB evaluation GPU.} At $K=1,2,4,8$, F1 is $1.000,1.000,1.000,0.944$, and exact-support rates are 100\%, 100\%, 100\%, and 80\%. {\color{black}At eight active sources, the guard/refinement/deletion/exchange modules raise exact recovery from the ablation's 50\% to 80\%.}

\begin{figure}[t]
\centering
\includegraphics[width=\columnwidth]{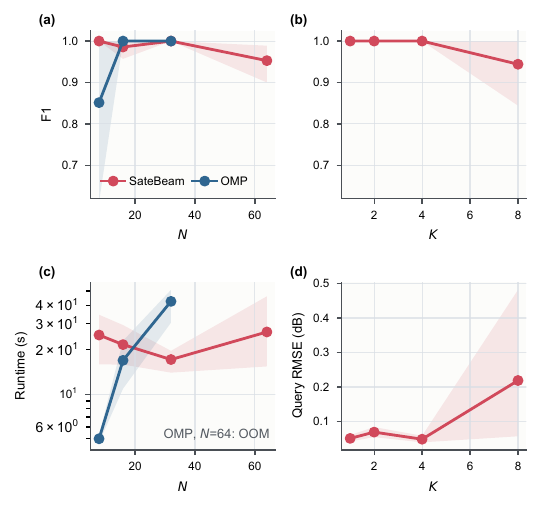}
\caption{Scaling with candidate count $N$ and active-set cardinality $K$; bands show 95\% CIs.}
\label{fig:scaling}
\end{figure}

\subsection{Order Selection and Support Audit}
\begin{figure}[t]
\centering
\includegraphics[width=\columnwidth]{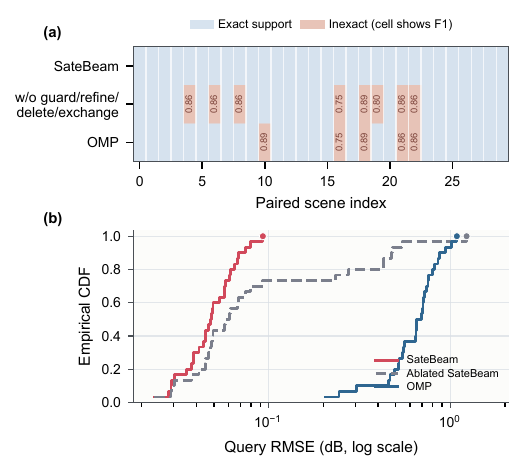}
\caption{Solver ablation and query-RMSE CDF over 30 paired scenes.}
\label{fig:support_audit}
\end{figure}

{\color{black}Under the $K=0$ null, BIC selects no source in all 30 independent scenes, for an observed family-wise false-positive rate of zero. Relative to the defined ablation, pre-BIC continuous refinement corrects five supports, backward deletion corrects two, and support exchange corrects one. Together these modules convert all eight ablation errors to exact recovery, and SateBeam is exact on all five scenes where OMP selects an incorrect support.} Figure~\ref{fig:support_audit}(a) reports every scene, and Fig.~\ref{fig:support_audit}(b) shows the full RMSE empirical distributions.

{\color{black}On a separate 20-scene $K=4$ rule-sensitivity set, standard BIC, AIC, and margin-BIC with $\tau=8$ all recover the exact support in every scene, with F1 $1.000$ and query RMSE 0.053 dB. Their median runtimes are 18.07, 17.88, and 18.30 s. These results establish consistent support and map estimates across the three declared penalties.} Standard BIC supplies the main model-order rule with penalty $4\log n$ per added source and zero additional margin.

\subsection{Empirical Identifiability}
The low- and high-coherence pairs have mean active--active coherence 0.021 and 0.551, and the strongest active--inactive coherence is approximately 0.79. {\color{black}With equal source powers, SateBeam attains F1/exact-support $1.000/100\%$ at low coherence and $0.955/85\%$ at high coherence, compared with OMP's $1.000/100\%$ and $0.905/75\%$.} {\color{black}With a 6 dB strength ratio, SateBeam attains $1.000/100\%$ at low coherence and $0.898/75\%$ at high coherence, compared with OMP's $0.720/20\%$ and $0.833/55\%$.} SateBeam achieves higher F1 and exact support in three groups and the same perfect result in the fourth. Its query RMSE values are 0.046, 0.122, 0.284, and 0.376 dB, versus OMP's 0.722, 0.818, 1.720, and 1.678 dB, establishing lower map error in every coherence--strength regime.

\section{Conclusion}
SateBeam exploits known ephemerides as complementary spatial and Doppler structure for joint active-satellite identification and radio-map reconstruction. The revised model exposes the link budget, circular-aperture beam, Doppler capture, Gamma energy statistics, and estimated background. The solver combines post-refinement BIC, backward deletion, and support exchange to deliver reliable source attribution and continuous beam recovery. {\color{black}In 30 paired real-ephemeris scenes, SateBeam achieves F1 $1.000$, 100\% exact support, and 0.050 dB independent-query RMSE, significantly outperforming OMP in F1 and map error while reducing median/P90 runtime from 15.60/34.18 s to 14.57/29.60 s relative to the guard/refinement/deletion/exchange ablation.} The mismatch and scaling experiments further establish consistent gains under independently generated sidelobes, elliptical beams, shadowing, blockage, 64 candidates, and eight active sources.
\bibliography{ref,satebeam_refs}
\bibliographystyle{IEEEtran}
\end{document}